\documentclass[twocolumn,preprintnumbers,amsmath,amssymb,aps,prb]{revtex4}
\usepackage{graphicx}
\begin{document}

\title{Dynamic Phases of Active Matter Systems with Quenched Disorder} 
\author{ Cs. S\'{a}ndor$^{1,2}$, A. Lib\'{a}l$^{1,2}$, 
C. Reichhardt$^{2}$ and C. J. Olson Reichhardt$^{2}$
}
\affiliation{$^{1}$Faculty of Mathematics and Computer Science, 
Babe\c{s}-Bolyai University, Cluj 400084 Romania}
\affiliation{$^{2}$Theoretical Division and Center for Nonlinear Studies,
Los Alamos National Laboratory, Los Alamos, New Mexico 87545, USA}
\date{\today}

\begin{abstract}
Depinning and nonequilibrium transitions within sliding states in systems 
driven over quenched disorder arise across a wide spectrum of
size scales ranging from atomic friction at the nanoscale,
flux motion in type-II superconductors at the mesoscale,
colloidal motion in disordered media at the microscale,
and
plate tectonics at geological length scales.
Here we show that active matter or 
self-propelled particles interacting with quenched disorder under an 
external drive represents a new class of system that can also exhibit
pinning-depinning phenomena, plastic flow phases, and nonequilibrium
sliding transitions that are correlated with distinct morphologies and
velocity-force curve signatures.
When interactions with the substrate are strong,
a homogeneous pinned liquid phase forms that depins plastically 
into a uniform disordered phase and then dynamically transitions first into a
moving stripe coexisting with a pinned liquid and then
into a moving phase separated state at higher drives.
We numerically map the resulting dynamical
phase diagrams as a function of external drive, substrate interaction strength,
and self-propulsion correlation length.
These phases can be observed for
active matter moving through random disorder.
Our results indicate
that intrinsically nonequilibrium systems can exhibit additional 
nonequilibrium  transitions when subjected to an external drive.  
\end{abstract}
\maketitle

\vskip2pc

\section{Introduction}

Depinning phenomena and dynamic phases of collective transport through
quenched disorder \cite{1,2} arise in a wide range of condensed matter systems
including
flux lines in superconductors \cite{3,4,5,6,7,8,9}, sliding charge density 
waves \cite{10}, moving electron crystals \cite{11,12}, magnetic skyrmions
\cite{13,14}, and driven pattern forming systems such as electronic states 
with competing interactions \cite{15}. In materials science systems such 
dynamics is relevant to sliding friction \cite{16}, dislocation motion
\cite{17}, yielding transitions \cite{18,19}, and models of fault lines 
and earthquakes \cite{20,21}. In soft matter, similar dynamics occurs in the 
depinning of 
contact lines \cite{22,23}, driven colloidal motion on disordered substrates
\cite{24,25,26}, or sliding of incommensurate colloidal structures on ordered substrates
\cite{27,28}.
Typically, these systems are in the pinned state at small
external drives, and as the drive increases, a transition to a moving 
state occurs at a specific critical value of the external force \cite{1,2}. 
At higher drives, distinct types of sliding 
motions can appear along with transitions between different dynamic states that can be 
deduced from features in the velocity-force curves.
The flow can be plastic 
or disordered \cite{1,2,3,4,5,8}, with the particles moving through riverlike 
channels\cite{24,25,26}, or elastic, where the particles maintain the 
same neighbors while moving \cite{1,3,24}. Transitions from disordered to more ordered or 
coherent flow can occur \cite{3,4,5,6,8,9}, such as from plastic flow to a moving 
smectic state \cite{6,7,8,9}.
Such transitions are associated
with changes in the moving structure morphologies, the density of topological 
defects \cite{4,8,9,14,24}, and the noise fluctuations \cite{5,8}.

In all of the systems mentioned above,
an external driving force produces a nonequilibrium condition.
Other types of
nonequilibrium  systems that involve no external driving force include
self-propelled or active matter systems.
These may be biological 
systems, such as swimming bacteria,
or artificial swimmers, such as 
self-propelled colloidal systems,
each of which
can exhibit
distinct nonequilibrium phases as a function of the activity or particle density
\cite{29,30}.
Many types of active matter systems can be effectively modeled 
as a collection of sterically interacting  disks undergoing driven Brownian 
diffusion \cite{31,32,33,34,35} or run and tumble dynamics \cite{35,36,37,38}.
These disks can form uniform liquid states
as well as phase separated states in which
dense disk clusters are separated by a low density disk gas.
An open question is whether such systems
would exhibit depinning transitions and different types 
of sliding states if an external drive were applied
when the disks are
coupled to a random substrate, as
the driven dynamics of particles 
interacting with obstacles generally are distinct from those of particles
on random pinning substrates.
Previous studies of active matter 
systems driven through obstacle arrays
showed that the average drift 
mobility varies non-monotonically with increasing activity, dropping when the 
system forms a phase-separated clump state \cite{38}.
Other studies of active matter involving swarming or Vicsek flocking models 
moving over disordered substrates in the absence 
of an external drive showed that there can be an optimal noise at which
flocking occurs \cite{39} as well as
transitions from flocking to non-flocking behavior \cite{40}. 

Here we use large scale GPU-based computer simulations to characterize the 
different states of run and tumble disks driven over a random pinning substrate,
and show that a rich variety of distinct dynamical phases are possible that can 
be identified by the morphologies of the moving structures as well as by features in 
bulk transport properties.
Despite the additional complexity introduced by
the self-propulsion, we find several generic features in the dynamic phase diagrams 
that are similar to those observed for non-active driven systems such as
superconducting vortex lattices, including
a disorder to order transition at higher drives.
In the 
limit of no quenched disorder, we find a cluster or phase-separated 
state, while in the presence of strong quenched disorder, the system forms a 
uniform pinned liquid state which undergoes plastic depinning as the drive is 
increased, followed at higher drives by a transition to a stripe state coexisting 
with the pinned liquid state.  At even higher drives, there is a transition to a
more fully phase separated state. We also find strong differences in the dynamic
phases for pinning arrays compared to those observed
in obstacle arrays, where
only a limited number of dynamic phases occur.

\section{Simulation}

We perform a simulation of $N = 5000$ to $N= 24000$ run and tumble disks
in a system with periodic 
boundary conditions of size $L \times L$ with $L=300$. The disks have a short range 
repulsive interaction modeled as a harmonic spring force 
${\bf F}_{\rm inter}=\Theta(d-2R)k(d-2R){\bf \hat{d}}$,
where $d$ is the distance between the centers 
of the disks, ${\bf \hat{d}}$ is the displacement vector between the disk centers,
$k=20.0$ is the spring constant, and $\Theta(x)$ is the Heaviside 
function. We use a disk radius of $R = 1.0$. The system density $\phi$ is 
defined by the area coverage of disks, $\phi=N \pi R^2/L^2$, giving us a range of 
$\phi = 0.1745$ to $\phi=0.8375$.  We randomly place $N_{p}$ 
non-overlapping pinning sites which are modeled as parabolic traps
that exert a force on the disks
of the form ${\bf F}_p =F_{p}(r/R_p)\Theta(r-R_p){\bf \hat{r}}$,
where ${\bf \hat{r}}$ is 
the displacement vector from the pinning site center to the disk center,
$r=|{\bf r}|$,
$F_{p}$ is the maximum pinning force, and $R_{p}$ is the pinning site radius.
We set $R_p=0.5$ to ensure that at most one disk can be pinned by a given pinning site.
We apply a uniform external drive ${\bf F}_{D}=F_D{\bf \hat x}$ on each disk in the 
$x$-direction and measure the resulting average drift velocity per 
disk in the direction of the drive 
$ \langle V \rangle = N^{-1} \sum^{N}_{i=0} \langle {\bf v}_{i} 
\cdot { \hat{ \bf{x} } } \rangle$ to produce velocity-force curves.

The disk dynamics is obtained
by integrating the following overdamped equation of motion:
$\eta {d{\bf r}}/{dt} = {\bf F}_{\rm inter} + {\bf F}_m + {\bf F}_p + {\bf F}_D$,          
where $\eta=1$ is the drag coefficient. The disk self-propulsion is produced by the
motor force ${\bf F}_{m}$ which acts on a given disk in a fixed
randomly chosen direction for a run time
$\tau$ that is chosen from a uniform random distribution
over the range $[t_r,2t_r]$.
After $\tau$ simulation time steps have elapsed,
the motor force instantly reorients to a new randomly chosen direction, which it
maintains for the next $\tau$ simulation time steps before reorienting again.
In the absence of other disks, pinning sites, or obstacles, a disk would move
a distance $D$ during one running time, where $D$ is uniformly randomly
distributed over the range $[r_l, 2r_l]$ with
$r_l=t_r F_m \delta t$.
We use $F_{m} = 1.0$ and run times ranging from $t_r=1000$ to
$t_r=2.4 \times 10^6$ 
simulation steps, with a simulation time step of $\delta t = 0.001$. To initialize 
the system, we place the disks in random configurations and
increase $F_D$ by increments of $\delta F_D=0.25$, spending $1 \times 10^6$
simulation time steps at each increment.

\begin{figure}
\includegraphics[width=3.5in]{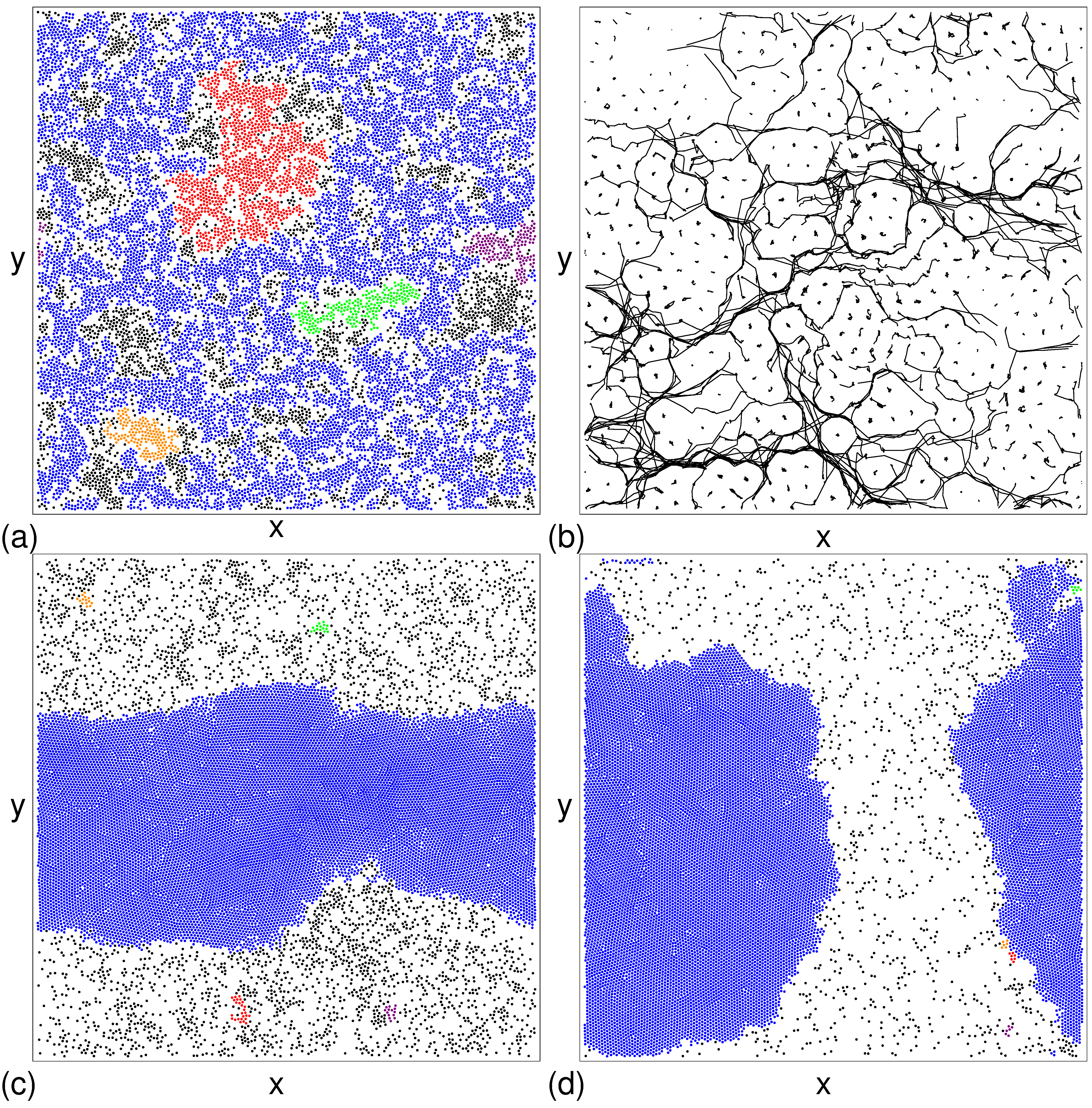}
\caption{ Images of disk locations (dots) in a system with
  $\phi = 0.55$,  $r_{l} = 300$, $N/N_{p} = 2.0$, and $F_{p} = 5.0$.
  Disks are colored according to the cluster
  to which they belong, with the largest cluster shown in blue, the second
  largest in red, the third largest in green, the fourth largest in purple, and the
  fifth largest in orange.
  (a) At $F_{D} = 0.0$, the system forms a uniform disordered state.
  (b) Disk trajectories (lines) in a portion of the sample at $F_{D} = 0.5$, 
  where the system is in a plastic flow state.
  (c) At $F_{D} = 3.25$, a dense stripe coexists with a pinned liquid.
  (d) At $F_{D} = 6.0$, a fully phase separated state appears.
}
\label{fig:1}
\end{figure}

\section{Results}

In Fig.~\ref{fig:1} we illustrate the disk locations
for a system with  $\phi = 0.55$, $r_{l}  = 300$, 
$N/N_{p} = 2.0$, and $F_{p} = 5.0$.
For these values of $\phi$ and $r_l$, in the absence of pinning
the system 
forms a phase separated state in which the disks
condense into a single large dense cluster.
When pinning is present, 
Fig.~\ref{fig:1}(a)
shows that at $F_{D}= 0$, the system forms a homogeneous disordered phase. 
We identify disks that are in contact with one another using the Luding and 
Herrmann cluster algorithm \cite{hermann}, and plot the largest cluster in blue, the second
largest cluster in red, and the next largest clusters in green, purple, and orange.
In Fig.~\ref{fig:1}(a) for $F_D=0$, the largest cluster
percolates across the entire sample.
At $F_{D} = 0.5$ in Fig.~\ref{fig:1}(b), we observe a plastic flow
state, as indicated by the riverlike disk trajectories, in which the overall disk
density remains uniform.
As $F_D$ increases,
a transition occurs into a moving stripe state as shown 
in Fig.~\ref{fig:1}(c) for $F_{D} = 3.25$.
Here a portion of the disks form a dense moving phase that
is aligned with the driving direction. The disks that are not in the dense phase
are directly trapped by the pinning sites.
As $F_{D}$ approaches $F_p$, more 
of the pinned disks become mobile and the system enters the fully phase separated 
state illustrated in Fig.~\ref{fig:1}(d) for $F_{D} = 6.0$,
where a single large clump appears
that has no particular orientation with respect to the driving direction.

\begin{figure}
\includegraphics[width=3.5in]{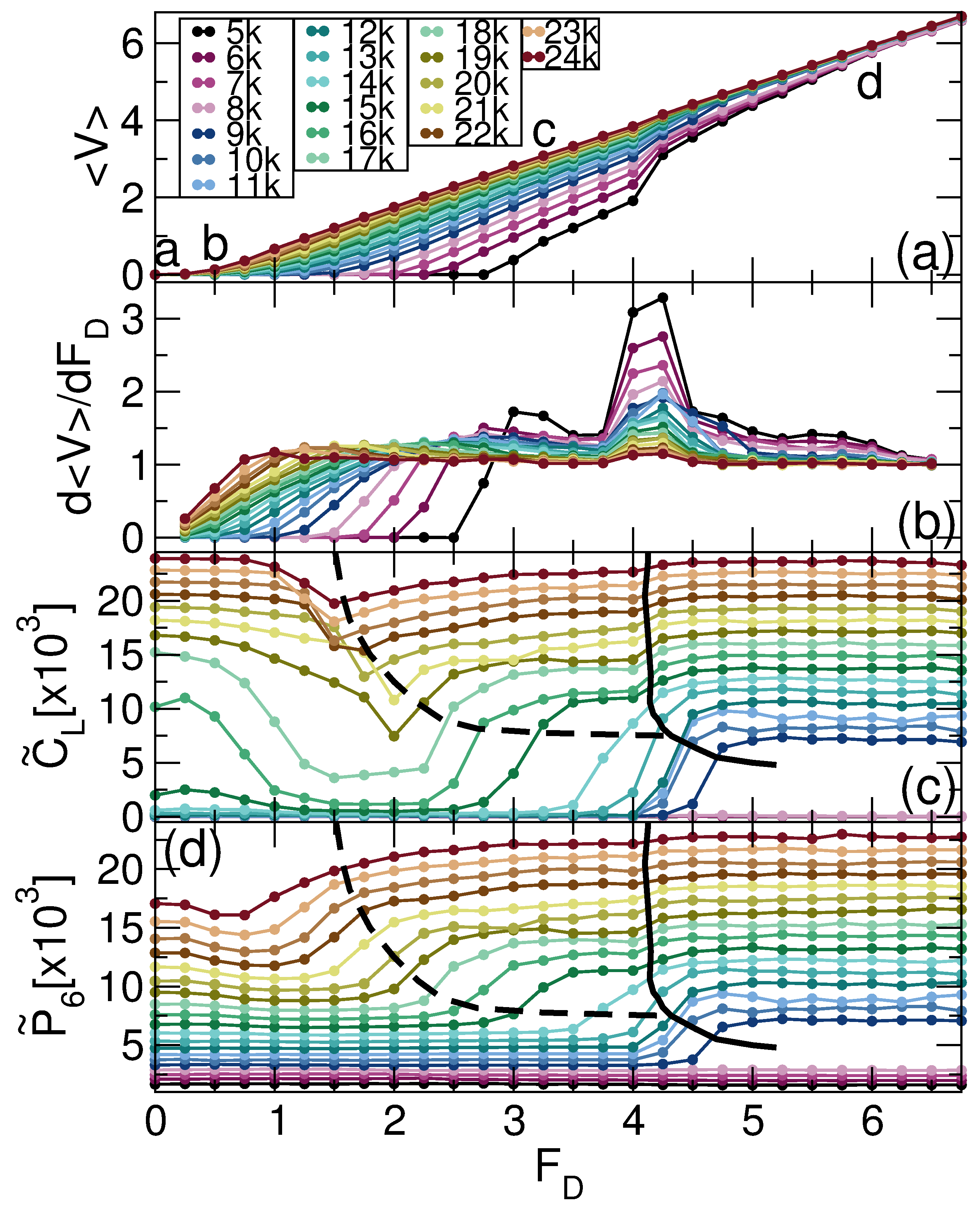}
\caption{(a) The velocity $\langle V\rangle$ vs $F_{D}$ for the system in
  Fig.~\ref{fig:1}(a) with
  $F_{p} = 5.0$, $r_l=300$,
  $N_{p} = 8000$, and varied $\phi$ ranging from $\phi = 0.17453$
  to $\phi=0.837$, corresponding to $N=5000$ to $N=24000$  in the
  figure legend.
  The letters {\bf a} to {\bf d} indicate
  the values of $F_{D}$ at which the images in
  Fig.~\ref{fig:1} were obtained for the $\phi = 0.55$ sample.
  (b) The corresponding $d\langle V\rangle/dF_{D}$
  vs $F_D$ curves.
  (c)
  The size of the largest cluster $\tilde{C}_{L}$ vs $F_{D}$.
The dashed line indicates the transition into the moving stripe state, and
  the solid line 
denotes the transition to the moving phase separated state.
  (d)
The number of six-fold coordinated disks $\tilde{P}_{6}$ vs $F_{D}$.
}
\label{fig:2}
\end{figure}

In Fig.~\ref{fig:2}(a) we plot the velocity-force curves for
the system in Fig.~\ref{fig:1} at disk densities ranging from
$\phi = 0.17453$ to $\phi=0.837$ in increments of
$\delta\phi=0.035$, while in Fig.~\ref{fig:2}(b) we show the 
corresponding $d\langle V\rangle/dF_{D}$ curves.
The letters {\bf a} to {\bf d} indicate 
the values of $F_{D}$ at which 
the images in Fig.~\ref{fig:1} were obtained for the
$\phi = 0.55$ sample,
demonstrating that the different phases correlate with distinct
features in the transport curves.
Previous studies of superconducting vortex systems showed that a
peak in $d\langle V\rangle/dF_D$, the derivative of the velocity-force curve, is
associated with a dynamical phase 
transition from plastic flow to a moving lattice state \cite{2,3,8}.
In more 
complex systems, such as driven pattern forming systems, multiple peaks in the 
$d\langle V\rangle/dF_D$ curves coincide with
multiple transitions in the structure of the moving state \cite{41,42}.
In Fig.~\ref{fig:2}(a),
a depinning transition occurs at a critical driving force
called $F_{c}$  
which shifts to lower values as $\phi$ increases.
The $d\langle V\rangle/dF_D$ curves 
in Fig.~\ref{fig:2}(b) have a double peak feature, with the largest peak near near 
$F_{D} = 4.0$ and a smaller peak at lower values of $F_{D}$.
Both 
peak features become less distinct as $\phi$ increases.
The transition from a pinned state to plastic flow occurs at the point at
which $d\langle V\rangle/dF_D$ first rises above zero, while
the moving 
stripe phase appears just above the first peak in
$d\langle V\rangle/dF_D$.
The second peak in $d\langle V\rangle/dF_D$
is associated with a transition to a moving fully phase 
separated state or to a uniformly dense state in which all the disks are in motion.

\begin{figure}
\includegraphics[width=3.5in]{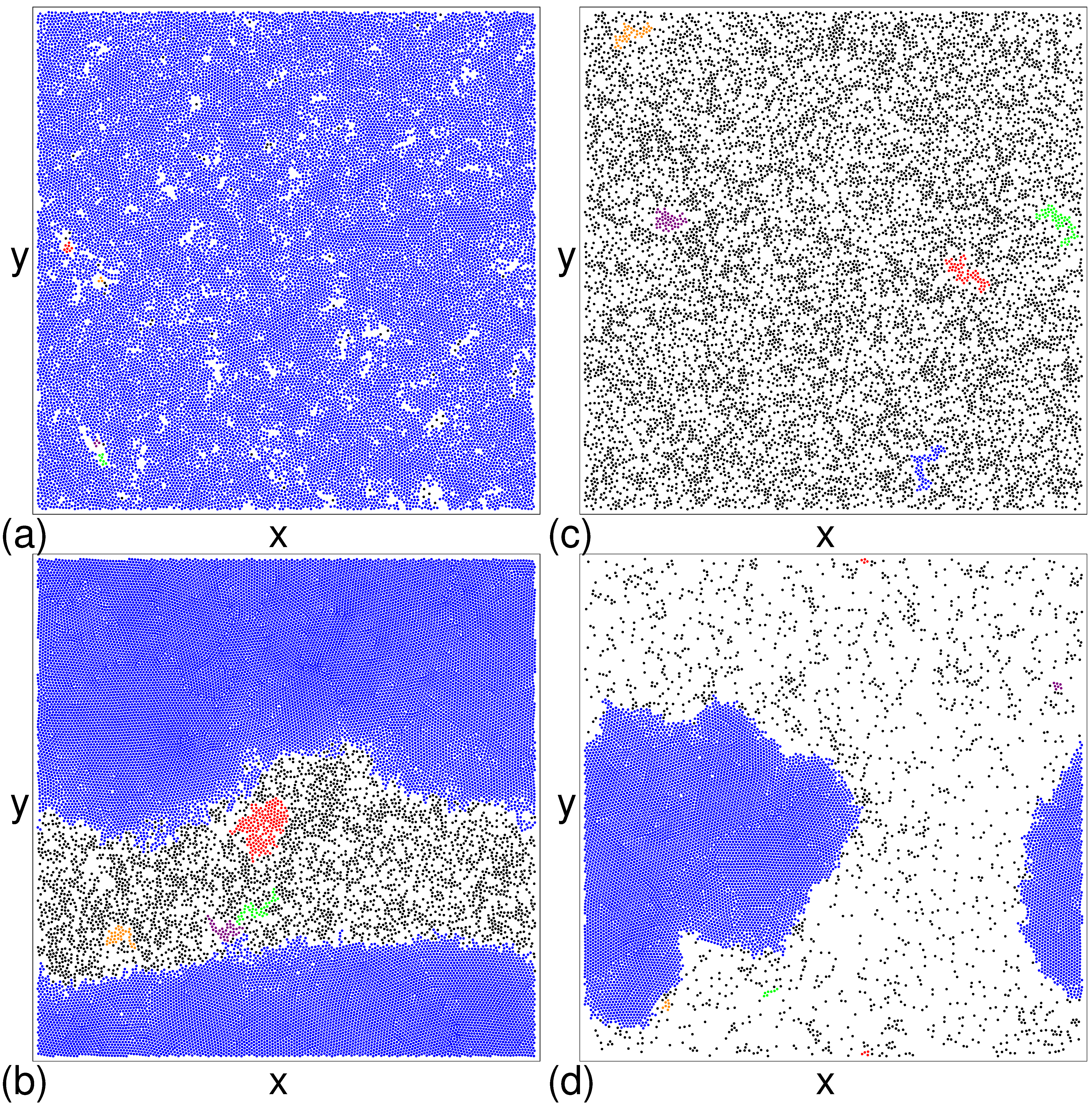}
\caption{Image of disk locations (dots) in a sample with
  $r_{l} = 300$, $N_{p} = 8000$, and $F_{p} = 5.0$.
  Disks are colored according to the cluster to which they belong, with the largest
  cluster shown in blue.
  (a) The uniform disordered state for $\phi=0.8375$ and $F_{D} = 0$.
  (b) The moving stripe state for $\phi=0.8375$ and $F_{D} = 1.5$.
  (c) The pinned liquid state for $\phi=0.35$ and $F_{D} = 0$.
  (d) The moving phase separated state for $\phi=0.35$ and $F_D=5.0$.
}
\label{fig:3}
\end{figure}

In Fig.~\ref{fig:2}(c,d)
we plot the size of the largest cluster $\tilde{C}_{L}$ and the number of six-fold 
coordinated disks $\tilde{P}_{6}$,
obtained from a Voronoi tessellation,
versus $F_{D}$ for the system in Fig.~\ref{fig:2}(a,b). For clarity, both $\tilde{C}_{L}$ 
and $\tilde{P}_{6}$ are plotted in terms of the total number of disks and
are not normalized to range between 0 and 1.
For $\phi < 0.314$, the system is in the moving disordered state for all values 
of $F_{D}$.
For $F_{D} = 0$, there is a pinned labyrinth phase
similar to that shown in Fig.~\ref{fig:1}(a)
for $\phi > 0.525$, while
for $\phi < 0.525$ there is
a pinned liquid state
similar to that illustrated in Fig.~\ref{fig:3}(c) at
$\phi = 0.35$.
At low drives, a drop in $\tilde{C}_L$ occurs
when the labyrinth phase breaks apart into a moving disordered phase,
as illustrated in Fig.~\ref{fig:1}(b).
The dashed line highlights an increase in $\tilde{C}_{L}$ that occurs when
the system enters a moving stripe phase.
In Fig.~\ref{fig:2}(c) the thick solid line highlights an increase in $\tilde{C}_{L}$ that occurs 
near $F_{D} = 4.0$, which is also the location of the largest peak in
$d\langle V\rangle/dF_D$.
In Fig.~\ref{fig:2}(d), there is an upward jump in $\tilde{P}_6$
near $F_{D} = 4.0$, marked with a solid line,
at the transition into the moving 
fully phase separated state.  At lower $F_D$ there is a smaller jump in $\tilde{P}_6$
marked with a dashed line connected with the transition into a moving stripe state.
There is an increase in $\tilde{P}_6$ whenever dense regions of disks form since
the densely packed regions have triangular ordering with disks that are mostly
sixfold coordinated.
We note that in the labyrinth phase,
$\tilde{C}_{L}$ can be large
since a cluster can percolate across the entire
system as illustrated in Fig.~\ref{fig:1}(a),
but
$\tilde{P}_{6}$  remains low
since the disks within the labyrinth are disordered.

In Fig.~\ref{fig:3}(a) we illustrate the pinned cluster state at $F_{D} = 0$ and 
$\phi = 0.8375$, while Fig.~\ref{fig:3}(b) shows the moving stripe state at
the same disk density for $F_{D} = 1.5$.
Figure~\ref{fig:3}(c) shows the pinned liquid phase at $F_{D} = 0$ and  
$\phi = 0.35$, where the clusters are absent,
and Fig.~\ref{fig:3}(d) illustrates the moving 
phase separated state at $\phi = 0.35$ and $F_{D} = 5.0$.

\begin{figure}
\includegraphics[width=3.5in]{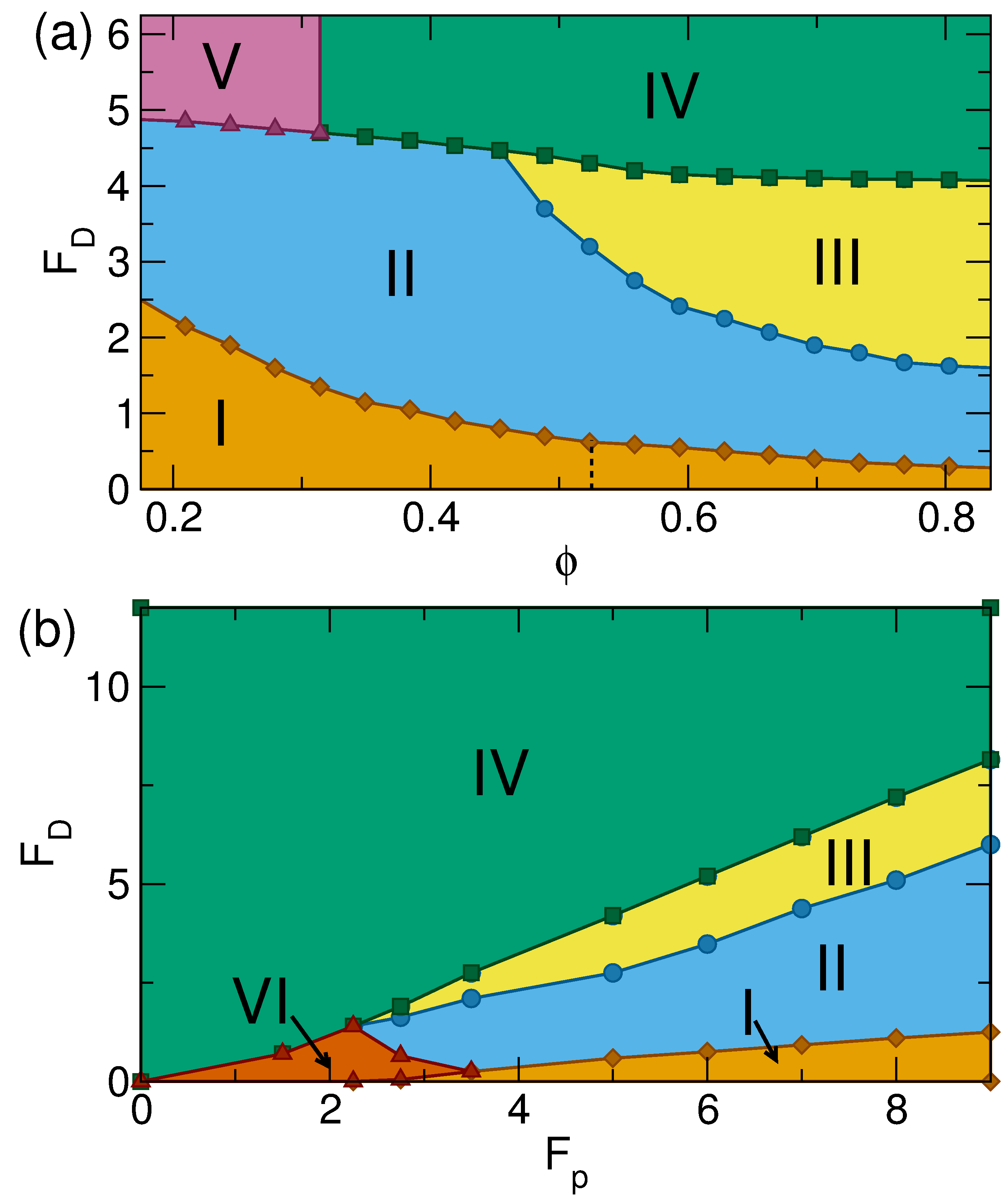}
\caption{
  (a) The dynamic phase diagram
  as a function of $F_D$ vs $\phi$ constructed from the features in Fig.~\ref{fig:2}
  for a sample with $F_{p} = 5.0$, $N_{p} = 8000$,
  and $r_l = 300$.
  I: pinned phase; II: plastic flow phase; III: moving stripe phase; IV: moving fully phase
  separated state; V: moving liquid phase.
  The dashed line in phase I indicates the separation between a pinned liquid and
  a pinned labyrinth state.
  (b) The dynamic phase diagram as a function of $F_{D}$ vs $F_{p}$
  for a sample with $\phi = 0.55$,
  $N/N_{p} = 2.0$, and $r_{l} = 300$.
  The phases I through V are marked as above.  Phase VI is
  the moving phase separated state in which some disks can be temporarily pinned.
}
\label{fig:4ab}
\end{figure}

\begin{figure}
\includegraphics[width=3.5in]{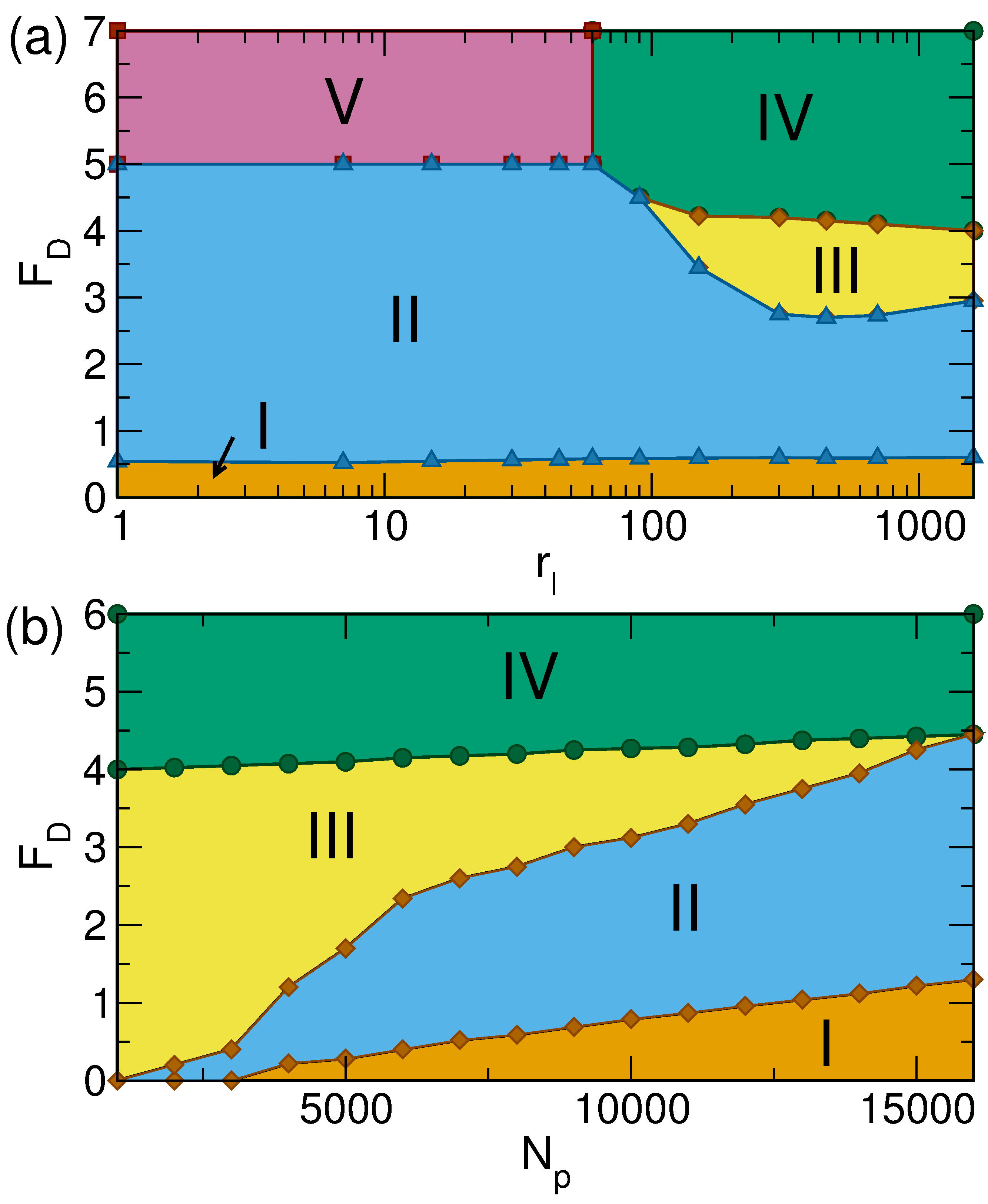}
\caption{
  (a) The dynamic phase diagram as a function of  $F_{D}$ vs $r_{l}$
  constructed from the features in Fig.~\ref{fig:2}
  for a sample with $\phi = 0.55$, $N_{p} = 8000$, and $F_{p} = 5.0$. 
  I: pinned phase; II: plastic flow phase; III: moving stripe phase; IV: moving fully phase
  separated state; V: moving liquid phase.
  (b) The dynamic phase diagram as a function of $F_{D}$ vs
  the number of pinning sites $N_{p}$ in samples with
$\phi = 0.55$, $F_{p} = 5.0$, and $r_{l} = 300$.  
}
\label{fig:4cd}
\end{figure}

From the features in $\tilde{C}_{L}$, $\tilde{P}_{6}$, and the transport curves in
Fig.~\ref{fig:2},
we identify various dynamic phases as plotted in 
Figs.~\ref{fig:4ab} and \ref{fig:4cd}.
As a function of $F_D$ versus $\phi$, shown in Fig.~\ref{fig:4ab}(a), we find
five dynamic phases, marked I through V.
There is a transition out of phase I,
the pinned phase,
at the depinning threshold $F_c$, which
drops to lower $F_D$ with increasing 
$\phi$.
A dashed line marks the transition between
$\phi < 0.525$, where the pinned clusters do not percolate and a pinned liquid
of the type shown in Fig.~\ref{fig:3}(c) forms,
and $\phi>0.525$,
where a pinned percolating cluster appears as illustrated
in Fig.~\ref{fig:1}(a) and Fig.~\ref{fig:3}(a).
In phase II, plastic flow of the type illustrated
in Fig.~\ref{fig:1}(b) occurs, and
the disk density is mostly homogeneous.
Phase III, the moving stripe phase,
is shown in Fig.~\ref{fig:1}(c) and Fig.~\ref{fig:3}(b).
Figures~\ref{fig:1}(d) and 3(d) illustrate phase IV, 
the moving fully phase separated
state, while in phase V, the moving liquid state, 
no clustering occurs but all the disks are moving.

\begin{figure}
\includegraphics[width=3.5in]{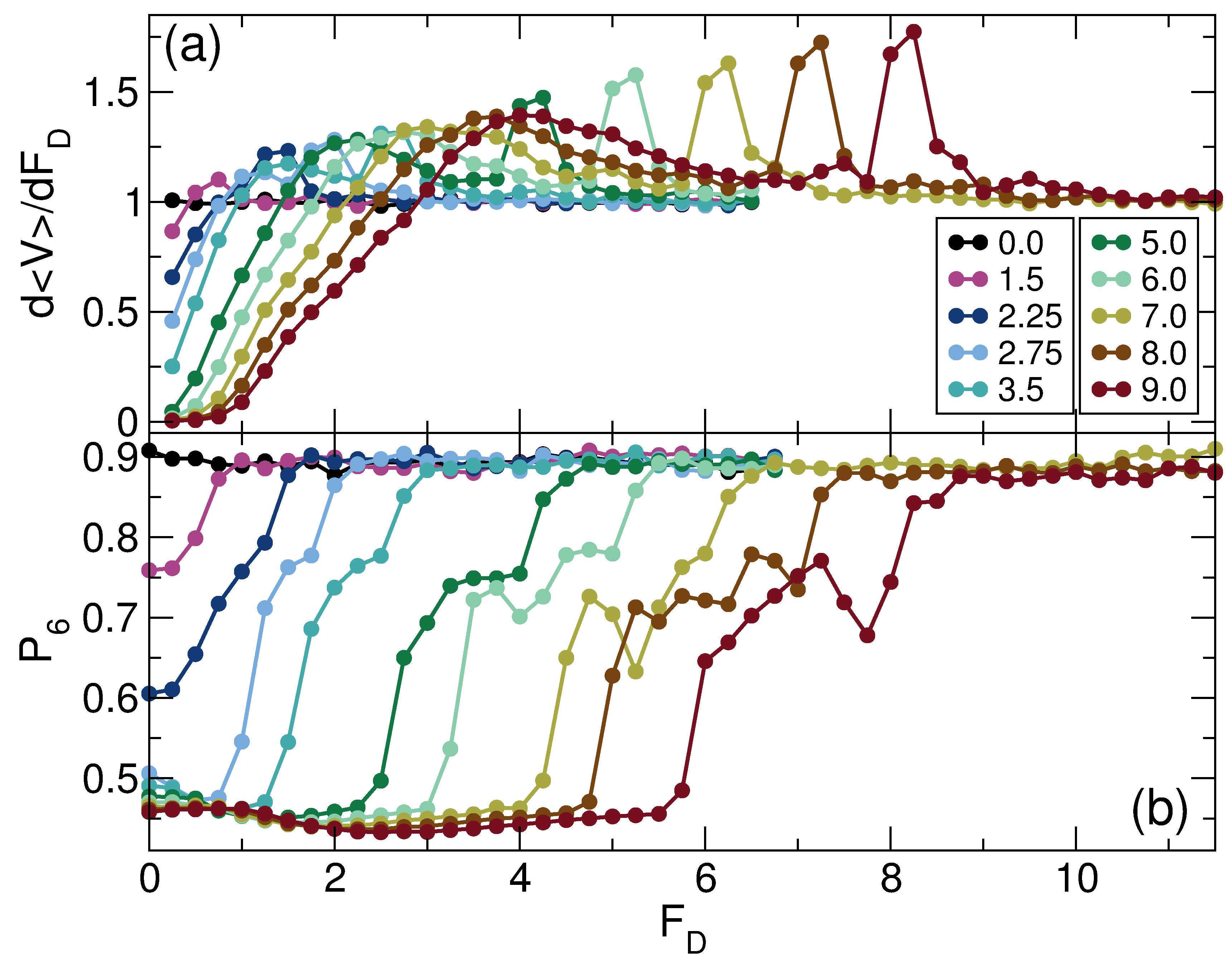}
\caption{ (a) $d\langle V\rangle/dF_{D}$ vs $F_D$ in samples with $\phi = 0.55$, 
  $N/N_{p} = 2.0$, and $r_{l} = 300$ from the system in Fig.~\ref{fig:4ab}(b)
  for different values of $F_p$ ranging from $F_p=0$ to $F_p=9.0$.
  As $F_p$ increases, a second peak appears and shifts to higher $F_D$.
(b) The corresponding $P_{6}$ vs $F_{D}$.
}
\label{fig:5ab}
\end{figure}

\begin{figure}
\includegraphics[width=3.5in]{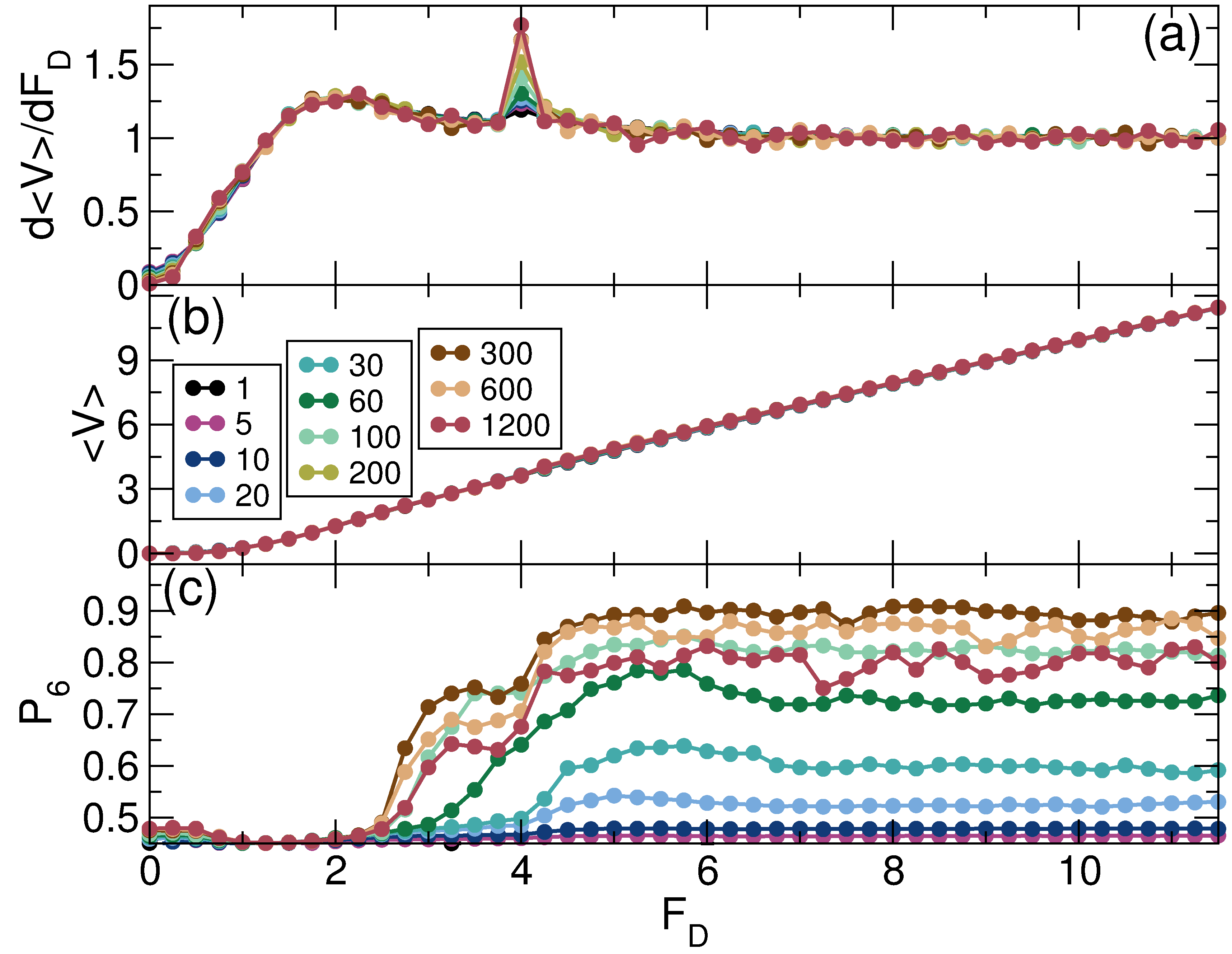}
\caption{
  (a)
  $d\langle V\rangle/dF_{D}$ vs $F_{D}$ for samples with
  $\phi = 0.55$, $F_{p} = 5.0$, and $N/N_{p} = 2.0$ for varied $r_{l}$
  ranging from $r_l=1$ to $r_l=1200$.
The magnitude of the second peak in $d\langle V\rangle/dF_{D}$ decreases with decreasing
$r_l$.
(b) The corresponding $\langle V\rangle$ vs $F_D$ curves.
(c) The corresponding $P_{6}$ vs $F_{D}$ curves,
indicating that
phase IV
disappears for $r_{l} < 60$.
}
\label{fig:5cde}
\end{figure}

We have also examined the dynamic phases as a function of
the pinning strength $F_p$.
In Fig.~\ref{fig:5ab}(a) we plot $d\langle V\rangle/dF_{D}$ versus $F_{D}$ 
for a system with $\phi = 0.55$ and $N/N_{p} = 2.0$
over the range $F_{p} = 0$ to 
$F_p=9.0$.
A double peak in $d\langle V\rangle/dF_D$ occurs only when $F_{p} > 1.5$, and 
both peaks shift to higher values of $F_{D}$ with increasing 
$F_{p}$.
Figure~\ref{fig:5ab}(b) shows the corresponding
normalized fraction of sixfold coordinated disks $P_{6}=\tilde{P}_6/N$ versus $F_{D}$. 
For $F_{p} > 3.5$, the disks are disordered and $P_{6} \approx 0.5$.
When the system enters phase III, a feature appears
near $P_{6} = 0.75$, 
and there is a jump up in $P_{6}$ at the onset of phase IV.
For low pinning strengths of
$F_{p} < 2.75$, a moving 
phase separated state called phase VI appears
in which some disks can be temporarily pinned, while
for large enough $F_D$ all the disks are moving and 
the system enters phase IV, the flowing state.
The dynamic phase diagram
as a function of $F_D$ versus $F_p$ in Figure~\ref{fig:4ab}(b)
indicates that phase III flow only occurs for $F_{p} > 2.0$, while phases I 
and III grow in extent with increasing $F_p$.

In Fig.~\ref{fig:5cde}
we plot the velocity-force and $d\langle V\rangle/dF_{D}$
curves for a system with $F_{p} = 5.0$, $\phi = 0.55$, 
and $N/N_{p} = 2$ for varied run lengths from
$r_l=1.0$ to $r_l=1200$.
The curves follow each other closely except for the second peak in 
$d\langle V\rangle/dF_{D}$, which increases in magnitude
with increasing $r_{l}$.
The second peak is absent at lower $r_l$ when a
transition from phase II to phase V occurs,
but materializes at larger $r_l$ once
a transition from phase III to phase IV
begins to occur.
In Fig.~\ref{fig:5cde}(c) we plot the corresponding $P_{6}$ versus $F_{D}$
curves which
show the onset of the transition into region IV in the form of an increase in $P_6$. 
In Fig.~\ref{fig:4cd}(a), the dynamic phase diagram
as a function of $F_D$ versus $r_l$ for the system in Fig.~\ref{fig:5cde}
shows that phases IV and III only occur for      
$r_{l} > 60$.
For $r_{l} < 60$, phase IV, the moving phase separated state, disappears,
and the system passes directly from
phase II plastic flow  to the phase V moving liquid.
Here, the
line marking the phase II to phase V transition remains flat at $F_D=F_p=5.0$.
We have also examined the case of fixed $\phi = 0.55$
and varied $r_l$
to obtain
the dynamic phase diagram as a function of $F_D$ versus $N_p$ in
Fig.~\ref{fig:4cd}(b).  Phase III diminishes in width until
$N_p/N = 1.0$.  For larger $N_p$,
there is a direct transition from phase II to phase IV, and phase III flow disappears.  
Our results indicate that the dynamic phases we observe persist over a wide range of
system parameters and represent generic features of this class of system.

\section{Dynamics in Obstacle Arrays}

\begin{figure}
\includegraphics[width=3.5in]{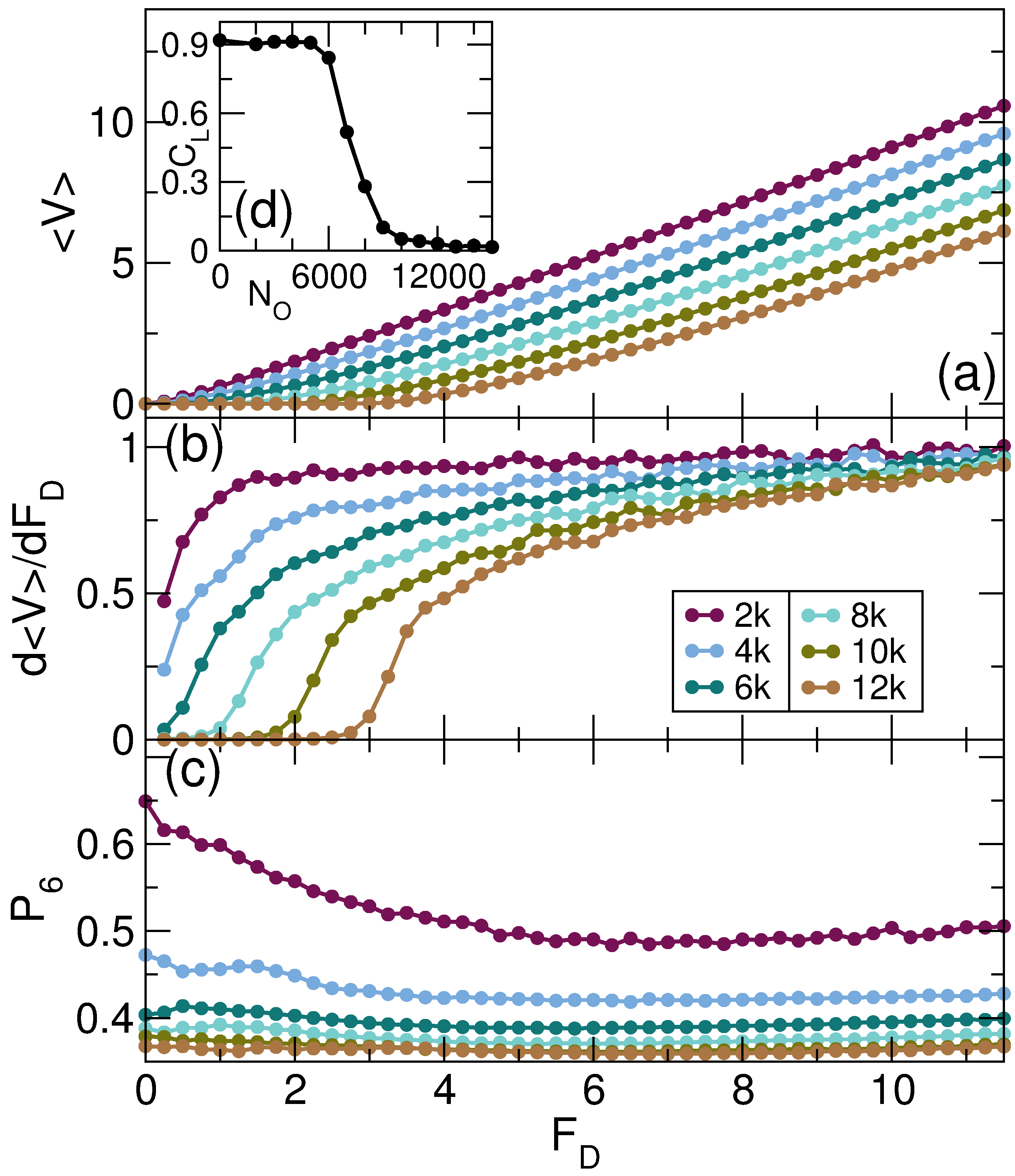}
\caption{
(a) $\langle V\rangle$ vs $F_{D}$ for systems with $\phi = 0.55$ and $r_{l} = 300$
for varied numbers of obstacles $N_o=2000$ to $N_o=12000$.
(b) $d\langle V\rangle/dF_{D}$ vs $F_{D}$ corresponding to the system in panel (a),
showing a lack of peak features.
(c) The corresponding normalized $P_{6}$ vs $F_{D}$ has a smooth monotonic behavior.
(d) The normalized cluster size $C_{L}$ vs number of obstacles $N_o$
for $\phi = 0.55$ and $r_{l} = 300$ at $F_{D} = 0.0$, showing a 
transition from a phase separated state
at low $F_D$ to a uniform state at high $F_D$.
}
\label{fig:6}
\end{figure}

\begin{figure}
\includegraphics[width=3.5in]{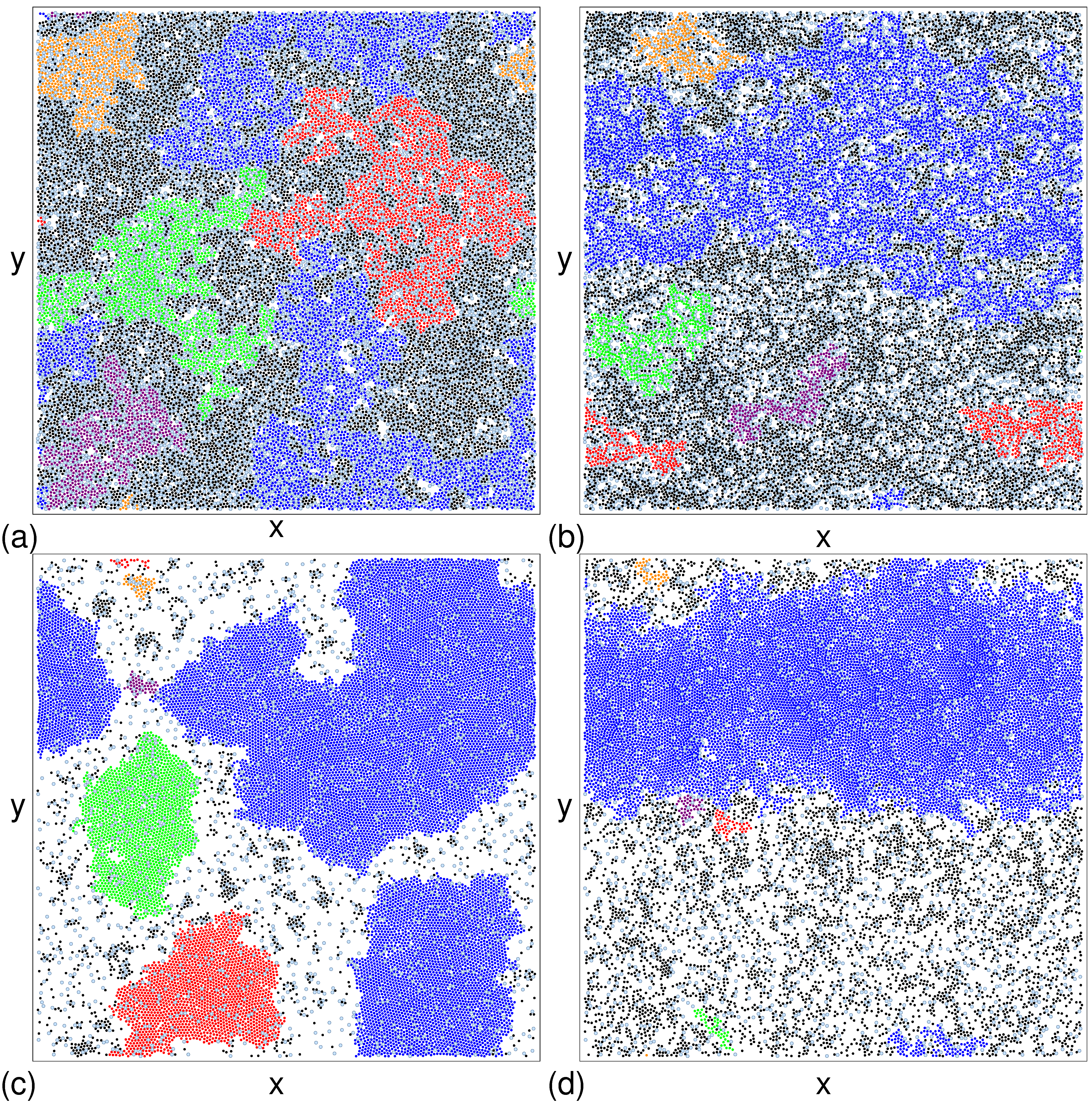}
\caption{Images of disk locations (dots) and obstacle locations (light blue circles)
  in samples with $\phi = 0.55$ and $r_{l} = 300$.
  (a) At $F_{D} = 0$ and $N_o=8000$ obstacles, a disordered state appears.
 (b) At $F_{D} = 6.0$ and $N_o=8000$, the system 
  is still disordered.
  (c) $F_{D} = 0$ and $N_o=2000$.
  (d) At $F_D=2.0$ and $N_o=2000$, a moving stripe state forms.}
\label{fig:7}
\end{figure}

Active disks moving through 
obstacle arrays have a very different behavior than that described above for
active disks in pinning arrays.
To explore this, we set up an obstacle landscape in which
the obstacles are modeled as immobile disks with a radius of $1.0$.
We consider a system with $r_l=300$, a mobile disk density
of $\phi = 0.55$, and $N_o$ obstacles.
An obstacle-free system at this disk density and running length $r_l$
forms a phase separated state for $F_{D} = 0$.
As we increase the number of obstacles $N_o$,
we find a transition
from a phase separated state to a uniform liquid state, as indicated in the plot of
the normalized value of $C_{L}$ versus $N_o$
in the inset of Fig.~\ref{fig:6}(a).
We show snapshots of the disk and obstacle positions in Fig.~\ref{fig:7}.  At $F_D=0$,
Fig.~\ref{fig:7}(a) indicates that a sample containing $N_o=8000$ obstacles
is disordered, while Fig.~\ref{fig:7}(c) shows that when $N_o=2000$ obstacles are
present, the system is density phase separated.
In Fig.~\ref{fig:6}(a,b,c) we plot
the velocity $\langle V\rangle$, $d\langle V\rangle/dF_{D}$, and $P_{6}$
versus $F_{D}$ for varied numbers of obstacles $N_o$.
As $N_o$ increases,
the average value of $\langle V\rangle$ monotonically deceases,
and the $d\langle V\rangle/dF_D$ curve contains no peaks, unlike the behavior for
a pinning substrate.
There is a gradual decrease in
$P_{6}$
with increasing $F_{D}$, and there 
are no peaks or dips in $P_6$ of the type that appear
for the pinning substrate.
In general, we find that the obstacles produce
no clear changes in the structure of the moving phase. 
When the number of obstacles $N_o>7000$, the system undergoes
plastic depinning into river-like channel flow, and remains in a uniform
density, disordered flow state up to
the highest drives $F_D$ that we considered, as shown in 
Fig.~\ref{fig:7}(b) for a sample with $N_o=8000$
at $F_{D} = 6.0$.
Conversely, for $N_o<7000$,
the system remains in a phase separated state, and forms a
moving stripe structure resembling that illustrated in Fig.~\ref{fig:7}(d)
for $N_o=2000$ at $F_{D} = 2.0$. This result 
indicates that there are no dynamical transitions in the moving state for
self-propelled disks moving through obstacle 
arrays; however, there is a transition in the drive-free limit of
$F_{D} = 0$ as a function of 
obstacle density.

We note that in course of completing this work we became aware
of an experimental paper by Morin
{\it et al.} \cite{New1}, who examined a colloidal
flocking active matter system drifting
through an obstacle array and found
that as the obstacle number increases,
the flow develops riverlike properties similar to that
observed in other depinning systems and
to what we find in Fig.~\ref{fig:1}(b).
In addition, Morin {\it et al.} report that there is
a critical obstacle density above which
the flocking behavior disappears,
not unlike the cluster disappearance at high obstacle densities that we observe
in Fig.~\ref{fig:7}(a).
Differences
between our system and that of Morin {\it et al.}
include the fact that the underlying physics behind
flock formation and cluster formation in the two systems is not the same;
additionally, Morin {\it et al.} considered only obstacles but not pinning sites.
The results of Ref.~\cite{New1} indicate that experiments on active matter
interacting with a substrate are feasible.
In addition to obstacle arrays, it is also possible to have active matter
move over landscapes of pinning sites created via optical means,
as has been demonstrated in other recent experiments \cite{New2,New3}.

\section{Summary}

We have shown that active matter assemblies driven over random pinning arrays 
represent a new class of system that exhibit pinning, depinning and 
nonequilibrium phase transitions similar to those found for driven vortices
and colloids moving over random disorder. In a regime where the system 
forms a phase separated state in the absence of pinning, we find that
the addition of pinning causes the formation of
a disordered uniform state that
depins plastically into a flowing uniform state with river-like features.
At higher drives, this is
followed by a transition to a moving dense stripe phase coexisting with a 
pinned liquid, until at the highest drives, the system transitions into a moving fully 
phase separated state. The different transitions are associated with
features in the velocity-force and
$d\langle V\rangle/dF_D$ curves that are similar to the features observed
in driven non-active systems with quenched disorder.
In contrast, for substrates composed of
obstacle arrays, there are no dynamical transitions in the moving state
and correspondingly there is 
a lack of features in the transport curves.
At zero drive, there is
a transition from  a phase separated 
state to a disordered state as the obstacle density increases.

\acknowledgments
This work was carried out under the auspices of the 
NNSA of the 
U.S. DoE
at 
LANL
under Contract No.
DE-AC52-06NA25396.
Cs. S\'{a}ndor and A. Lib\'{a}l
thank the Nvidia Corporation 
for their graphical card donation that was used in carrying out 
these simulations.

\end{document}